\documentclass[10pt,twocolumn]{aastex63}

\usepackage{chngcntr}
\usepackage{upgreek}

\newcommand{\cmst}{$\rm cm\,s^{-2}$}
\newcommand{\mst}{$\rm m\,s^{-2}$}

\newcommand{\logg}{log\,$g$ }
\newcommand{\teff}{$T_{\rm eff}$}

\received{May 12, 2021}
\revised{August 11, 2021}
\accepted{\today}
\submitjournal{ApJ}

\shorttitle{Water Latent Heat on Ultra-Cool Objects}
\shortauthors{Tang et al.}

\begin{document}

\title{Impacts of Water Latent Heat on the Thermal Structure of Ultra-Cool Objects: \\
        Brown Dwarfs and Free-Floating Planets}

%-----------------------------------
\author[0000-0003-4247-1401]{Shih-Yun Tang}
    \affiliation{Lowell Observatory, 1400 West Mars Hill Road, Flagstaff, AZ 86001, USA}
    \affiliation{Department of Astronomy and Planetary Science, Northern Arizona University, Box 6010, Flagstaff, AZ 86011, USA}

\author[0000-0002-3196-414X]{Tyler D. Robinson}
    \affiliation{Department of Astronomy and Planetary Science, Northern Arizona University, Box 6010, Flagstaff, AZ 86011, USA}
    \affiliation{Habitability, Atmospheres, and Biosignatures Laboratory, Northern Arizona University, Flagstaff, AZ 86011, USA}
    \affiliation{NASA Astrobiology Institute’s Virtual Planetary Laboratory, University of Washington, Box 351580, Seattle, WA 98195, USA}

\author[0000-0002-5251-2943]{Mark S. Marley}
    \affiliation{Department of Planetary Sciences and Lunar and Planetary Laboratory, The University of Arizona, 1629 University Blvd., Tucson, AZ 85721, USA}

\author[0000-0003-1240-6844]{Natasha E. Batalha}
    \affiliation{Space Sciences Division, NASA Ames Research Center, Moffett Field, CA 94035, USA}
 
\author[0000-0003-3444-5908]{Roxana Lupu}
    \affiliation{BAER Institute, NASA Ames Research Center, Naval Air Station, Moffett Field, CA 94035, USA}

\author[0000-0001-7998-226X]{L. Prato}
    \affiliation{Lowell Observatory, 1400 West Mars Hill Road, Flagstaff, AZ 86001, USA}
    \affiliation{Department of Astronomy and Planetary Science, Northern Arizona University, Box 6010, Flagstaff, AZ 86011, USA}

\email{sytang@lowell.edu}
%-----------------------------------
%-----------------------------------

\begin{abstract}

Brown dwarfs are essential targets for understanding planetary and sub-stellar atmospheres across a wide range of thermal and chemical conditions. As surveys continue to probe ever deeper, and as observing capabilities continue to improve, the number of known Y dwarfs\,---\,the coldest class of sub-stellar objects, with effective temperatures below about 600\,K\,---\,is rapidly growing. Critically, this class of ultra-cool objects has atmospheric conditions that overlap with Solar System worlds and, as a result, tools and ideas developed from studying Earth, Jupiter, Saturn and other nearby worlds are well-suited for application to sub-stellar atmospheres. To that end, we developed a one-dimensional (vertical) atmospheric structure model for ultra-cool objects that includes moist adiabatic convection, as this is an important process for many Solar System planets. Application of this model across a range of effective temperatures (350, 300, 250, 200\,K), metallicities ([M/H] of 0.0, 0.5, 0.7, 1.5), and gravities (\logg of 4.0, 4.5, 4.7, 5.0) demonstrates strong impacts of water latent heat release on simulated temperature-pressure profiles. At the highest metallicities, water vapor mixing ratios reach an Earth-like 3\%, with associated major alterations to the thermal structure in the atmospheric regions where water condenses. Spectroscopic and photometric signatures of metallicity and moist convection should be readily detectable at near- and mid-infrared wavelengths, especially with {\it James Webb Space Telescope} observations, and can help indicate the formation history of an object.

\end{abstract}

%-----------------------------------
\keywords{brown dwarf -- planets and satellites: atmospheres -- planets and satellites: detection -- stars: atmospheres}
%---------------------------------------------------------------
%%%%%%%%%%%%%%%%%%%%%%%%%%%%%%%%%%%%%%%%%%%%%%%%%%%%%%%%%%%%%%%%%%%%%%%%%%
%---------------------------------------------------------------
\section{Introduction} \label{sec:intro}

Brown dwarfs are important and abundant worlds with properties that span from stars to giant planets \citep{tarter1986}. Unlike stars, brown dwarfs are not massive enough to sustain core hydrogen fusion \citep{kumar1963}. Thus, brown dwarfs simply cool over cosmic timescales following their formation \citep{burrowsetal1997,chabrieretal2000,baraffeetal2002}.

As a given brown dwarf cools, and depending on its initial mass, its atmosphere can pass through a range of conditions appropriate for the formation of condensates. Spanning the effective temperature range of L dwarfs (roughly 2,400--1,400\,K), materials like corundum, iron, and silicates can condense to form optically thick clouds \citep{stevenson1986,lunineetal1989,burrows&sharp99,marleyetal2002,visscheretal2010,marleyetal2013}. Then for T dwarfs (with effective temperatures spanning roughly 1,400--600\,K), iron and rock particulate clouds clear from the photosphere \citep{saumon&marley2008,marleyetal2010} and proposed sulfide and salt species condense \citep{lodd99,morleyetal2014}. Finally, in Y dwarfs at the coolest end of the brown dwarf sequence (with effective temperatures below about 600\,K), water clouds are expected to condense in the atmosphere \citep{burrowsetal2003,morleyetal2014}. 

While the theory of a condensate sequence for sub-stellar objects is relatively well-developed, associated impacts on atmospheric structure and emergent spectra\,---\,beyond the influences of cloud opacity\,---\,have received extremely limited attention. Here, especially, the role of latent heat release during condensation is virtually unstudied. While the atmospheric abundances of the vapor-phase species that give rise to iron, rock, sulfide, and salt clouds are generally small enough \citep[{$\lesssim 10^{-6}$};][]{burrows&sharp99} such that associated latent heating can be neglected, this will not be the case for water in Y dwarf atmospheres. For solar metallicity Y dwarfs, atmospheric water vapor can approach volume mixing ratios of 0.1\% which, when paired with the relatively large latent heat of water ($\sim 40.7$ kJ mole$^{-1}$), implies moist adiabatic adjustments to the convective lower atmosphere \citep{Li18} at the 10\% level. Such adjustments would be even more substantial at super-solar metallicities.

Moist convection is well-studied in the Solar System, especially for Earth \citep{stevens2005}. Early studies of moist convection on other Solar System bodies \citep{barcilon&gierasch1970,gierasch1976,stoker1986,lunine&hunten1987} emphasized water in the atmosphere of Jupiter, the onset of condensation, and impacts on inferred oxygen abundance. Since that time, simulations and observations have explored moist convection and convective storms across the Solar System \citep{hueso&sanchezlavega2006,lian&showman2010}, especially for Jupiter and Saturn \citep{ingersolletal2000,hueso&sanchezlavega2001,hueso&sanchezlavega2004,sugiyamaetal2014,li&ingersoll2015}.

Fundamental work on Y dwarfs \citep{burrowsetal2003,morleyetal2014} stresses the importance of water clouds, but stops short of discussing the moist convection that could drive the formation of these clouds. \citet{tan17} used a three-dimensional general circulation model to explore how latent heat release from enstatite condensation might impact atmospheric circulation of T dwarfs. These first-of-their-kind simulations showed limited storm activity stemming from ``moist'' enstatite convection (likely influenced by relatively small vapor phase mixing ratios) and did not generate spectral observables for comparisons to observations. Finally, recent studies \citep{tremblinetal2015,tremblinetal2016} have explored simulated L- and T-dwarf atmospheres with prescribed reductions in the adiabatic temperature gradient, partially motivated by convection across the boundaries that define thermochemical transitions between CO/CH$_4$ and N$_2$/NH$_3$. While not highlighted in recent work, latent heating effects are another potential mechanism for reducing temperature gradients.

Work presented here explores the effect of water latent heat release on the atmospheric thermal structure and the emergent spectra of ultra-cool objects. Simulations span a range of gravitational accelerations and atmospheric metallicities that apply to Y-class brown dwarfs and free-floating planets. As metal enrichment influences water abundance and, in turn, the water vapor mixing ratio impacts latent heating effects, constraints on the thermal structure of ultra-cool objects may yield insights into their formation history. A prediction of the core accretion model for planet formation \citep{poll96} is that gaseous planets should demonstrate increasing metal enrichment with decreasing mass, whereas brown dwarfs formed from gravitational collapse would show metallicities akin to those of stars \citep[see Figure~8 from][]{zhang2020}. As an example of the scales involved, the atmospheres of Uranus and Neptune are enriched in carbon by roughly 100 times relative to solar (i.e., [C/H] of 2), while self-luminous objects in the solar neighborhood have an [Fe/H] that spans roughly $-0.7$ to 0.5 \citep{bude19}.  Nevertheless, recent inferences (from low signal to noise observations) show that Jupiter-mass exoplanets may possess metallicities spanning sub-Jupiter values to enhancements potentially greater than the Solar System ice giants \citep{zhang2020,Welbanks19}.

In what follows, Section~\ref{sec:model} describes a one-dimensional (vertical) radiative-convective model for sub-stellar objects that includes latent heating effects. Then, Section~\ref{sec:results} presents equilibrium thermal structures generated from this radiative-convective model as well as associated spectra, photometry, and detectability studies for {\it James Webb Space Telescope} ({\it JWST}). Finally, Sections~\ref{sec:disc} and \ref{sec:conc} discuss the implications of the models developed here and summarize key conclusions.

%---------------------------------------------------------------
%%%%%%%%%%%%%%%%%%%%%%%%%%%%%%%%%%%%%%%%%%%%%%%%%%%%%%%%%%%%%%%%%%%%%%%%%%
%---------------------------------------------------------------
\section{Model Description} \label{sec:model}

The one-dimensional (vertical) atmospheric structure model used in this study is derived from original brown dwarf simulation efforts described in \citet{marleyetal1996} and a Titan radiative-convective modeling tool developed by \citet{mckayetal1989}. Building on this foundational work, the one-dimensional model has seen extended applications to exoplanets \citep[e.g.,][]{fortneyetal2005,fortneyetal2008,morleyetal2017} and brown dwarfs \citep[e.g.,][]{marleyetal1996,saumon&marley2008,morleyetal2012,morleyetal2014,robinson&marley2014,marl21}. Most fundamentally, the model generates atmospheric thermal structure profiles that are in radiative-convective equilibrium and that assume local rain-out thermochemical equilibrium. 

Radiative fluxes are computed via the ``two-stream source function'' technique \citep{toon89} with the opacity databases from \citet{free08,free14} that have spectrally-resolved gas opacities incorporated via eight-term correlated-$k$ coefficients \citep{goodyetal1989,lacis&oinas1991}.
Clouds, when included, are modeled according to the {\tt EddySed} framework described in \citet{ackerman&marley2001}. Finally, chemical abundances of all radiatively-active species are based on  \citet{lodders&fegley2002,lodd04,lodd06,visscheretal2006,visscheretal2010}, and are generally a function of pressure, temperature, and metallicity ([M/H]). Metallicity is taken as the logarithm of the metal (i.e., atomic species heavier than helium) enrichment relative to solar abundance, then, quantifies the multiplicative enhancement of all metals in the assumed underlying atomic abundances in the thermochemical calculation. For a recent review of the solar elemental composition, see \citet{lodders2020}. 

%---
\subsection{Convection}\label{sec:convect}

Modeling results presented here adopt the timestepping framework outlined in \citet{robinson&marley2014}, which has been found to produce better-converged stratospheric temperature profiles compared to a Newton-Raphson solver. The convective heat flux, $F_{\rm c}$, is computed using mixing length theory \citep{prandtl1925,vitense1953,bohmvitense1958,gierasch&goody1968} with
\begin{equation} \label{eq:fc}
    F_{\rm c} = -\rho c_{\rm p} K_{\rm H} \cdot \frac{T}{\ell} \cdot \left( \nabla_{{\rm ad}} - \nabla \right)
\end{equation}
where $\rho$ is the atmospheric mass density, $c_{\rm p}$ is the gas-phase, temperature-dependent specific heat capacity, $K_{\rm H}$ is the eddy diffusivity for heat, $T$ is  temperature, $\ell$ is the mixing length (often taken to scale with the atmospheric pressure scale height, $H_{\rm p}$), $\nabla_{{\rm ad}}$ is the isentropic adiabatic temperature gradient, and $\nabla = \partial \ln T/\partial \ln p$ (with $p$ as atmospheric pressure). The eddy diffusivity is given by,
\begin{equation}
    K_{\rm H} = \ell^{2} \sqrt{ \frac{g}{H_{\rm p}}  \left( \nabla - \nabla_{\rm ad} \right) }\ ,
\end{equation}
where $g$ is the gravitational acceleration. A convective flux is only applied when the atmosphere is unstable to convection (i.e., when $\nabla > \nabla_{\rm ad}$); the eddy diffusivity and convective flux are zero in all stable portions of the atmosphere.

Equilibrium thermal structure solutions that include latent heating effects are emphasized in this work. These ``moist'' models relax the convective portions of the atmosphere to the moist pseudo-adiabat with,
\begin{equation}\label{eq:nabla_m}
    \nabla_{\rm m} = \frac{1 + f\frac{L}{R_{\rm U}T}}{\nabla_{\rm d}^{-1} + f\frac{L^2}{R_{\rm U}^2 T^2} } \ ,
\end{equation}
where $f$ is the gas phase volume mixing ratio for the condensing species, $L$ is the latent heat, $R_{\rm U}$ is the universal gas constant, and $\nabla_{\rm d}$ is the adiabat for the ``dry'' portion of the atmosphere \citep[for a complete derivation of the isentropic moist adiabat, see][their Section~2]{Li18}. While the models presented here adopt dry adiabats from the equation of state work of \citet{saumonetal1995}, the pressure and temperature conditions are generally such that an ideal gas equation of state would suffice. A distinct treatment that would more properly capture the dynamic evolution of the atmosphere would only convectively mix the atmosphere when the local temperature gradient is steeper than the dry adiabatic gradient, and then release latent heat to the atmosphere if the mixed layers experience subsequent condensation. Note that the ``pseudo'' in the pseudo-adiabat refers to the assumption that the condensed phase leaves the parcel, which is consistent with our assumption of local rain-out thermochemical equilibrium. Finally, ``dry'' models\,---\,which do not consider latent heating effects on the adiabat\,---\,simply adopt $\nabla_{\rm d}$ as the adiabat within the mixing length formalism.

%---
\subsection{Timestepping Solutions}

The timestepping model was initialized with a first estimate from a Newton-Raphson solver. Individual timesteps were permitted to adapt in duration to ensure that the Courant–Friedrichs–Lewy condition is met. Within each timestep ($\Delta t$), atmospheric temperatures are updated according to
\begin{equation} \label{eq:ts}
    T_i(t+\Delta t) = T_i(t)+Q_i \Delta t
\end{equation}
where Q$_i$ is the heating rate at $i^{\rm th}$ pressure level given by
\begin{equation} \label{eq:q}
    Q = \frac{\partial T}{\partial t} = \frac{g}{c_{\rm p}} \frac{\partial F_{\rm net}}{\partial p}
\end{equation}
and $F_{\rm net}$ is the net energy flux (i.e., the sum of the net convective and radiative fluxes). Models are considered converged once the net flux at each model level is within a fraction of $10^{-4}$ of the internal energy flux (i.e., $\sigma T_{\rm eff}^4$, where $\sigma$ is the Stefan-Boltzmann constant and $T_{\rm eff}$ is the adopted effective temperature of the sub-stellar object).

%---
\subsection{Model Grid}

Table~\ref{tab:gridpar} shows the modeling parameter grid adopted in this study. For this initial work, models are assumed to be cloud-free except for a limited number of example partially-clouded cases with \logg of 4.0, a moist adiabatic troposphere, \teff{} of either 200 or 250\,K, and [M/H] of either 0.0 or 1.5. The grid parameters were primarily selected to explore scenarios where water latent heat effects may be relevant to the atmospheres of ultra-cool objects. The minimum \teff{} in the grid is slightly cooler, by 50K, than the \teff{} of the coolest Y dwarf known to date, WISE 0855$-$0714 \citep[$\sim 250$\,K;][]{luhm14}. Also, the maximum metallicity is taken to be intermediate to Uranus/Neptune (with [M/H] $\sim$ 2) and Saturn (with [M/H] $\sim$ 1) \citep[see~Table~1 in][]{atre20}. 
For clarity, we label objects with stellar-like metallicities ([M/H]$\lesssim$0.5) as ``Y-class brown dwarfs'' (YBs), and label objects with super-stellar metallicities ([M/H]$\gtrsim$0.5) as ``Y-class planets'' (YPs)\footnote{Our labeling is primarily for convenience and clarity, and we recognize that planets can have [M/H] $<$ 0.5.}.

The estimated mass range for worlds explored in this study is about 0.004--0.005\,$M_\sun$ ($\sim$4.3--5.0 $M_{\rm Jupiter}$) for \logg of 4.0 and about 0.026--0.028\,$M_\sun$ ($\sim$27.2--29.3 $M_{\rm Jupiter}$) for \logg of 5.0, based on cloud-free evolutionary models from the \texttt{Sonora Bobcat} grid \citep{marl21}.
As noted in \citet{morleyetal2014}, a square grid (as is adopted here) can include some parameter combinations (e.g., of effective temperature and gravity) that may be unphysical given the mass-dependent cooling models of \citet{saumon&marley2008}. For some higher \logg cases presented here, evolutionary models indicate that it may take tens of Gyrs for such worlds to cool to ultra-cool temperatures.

%-----------------------------------
\startlongtable
\begin{deluxetable}{c cl}
\tablecaption{Modeling Grid\label{tab:gridpar}
              }
\tabletypesize{\footnotesize}
\tablehead{ 
 	 \colhead{Parameter}    & \colhead{Units}    & \colhead{Values}
 	 }
\startdata
Gravity$^\dagger$               & \mst  & 100, 300, 500, 1000       \\
Effective temperature (\teff)   & K     & 200, 250, 300, 350        \\
Adiabat option                  &       & dry, moist                \\
{Metallicity ([M/H])}           &       & YBs$^\ddagger$: \{0.0, 0.5\},         \\
{                   }           &       & YPs$^\ddagger$: \{0.7, 1.0, 1.5\}     
\enddata
\tablecomments{
    $^\dagger$Gives \logg{} of roughly 4.0, 4.5, 4.7, 5.0 (in \cmst). 
    $^\ddagger$ YBs: ``Y-class brown dwarfs''; YPs: ``Y-class planets''.
    }
\end{deluxetable}

%---------------------------------------------------------------
%%%%%%%%%%%%%%%%%%%%%%%%%%%%%%%%%%%%%%%%%%%%%%%%%%%%%%%%%%%%%%%%%%%%%%%%%%
%---------------------------------------------------------------
%% fig: PT-profile
\begin{figure*}[tb!]
\centering
\includegraphics[angle=0, width=1.\textwidth]{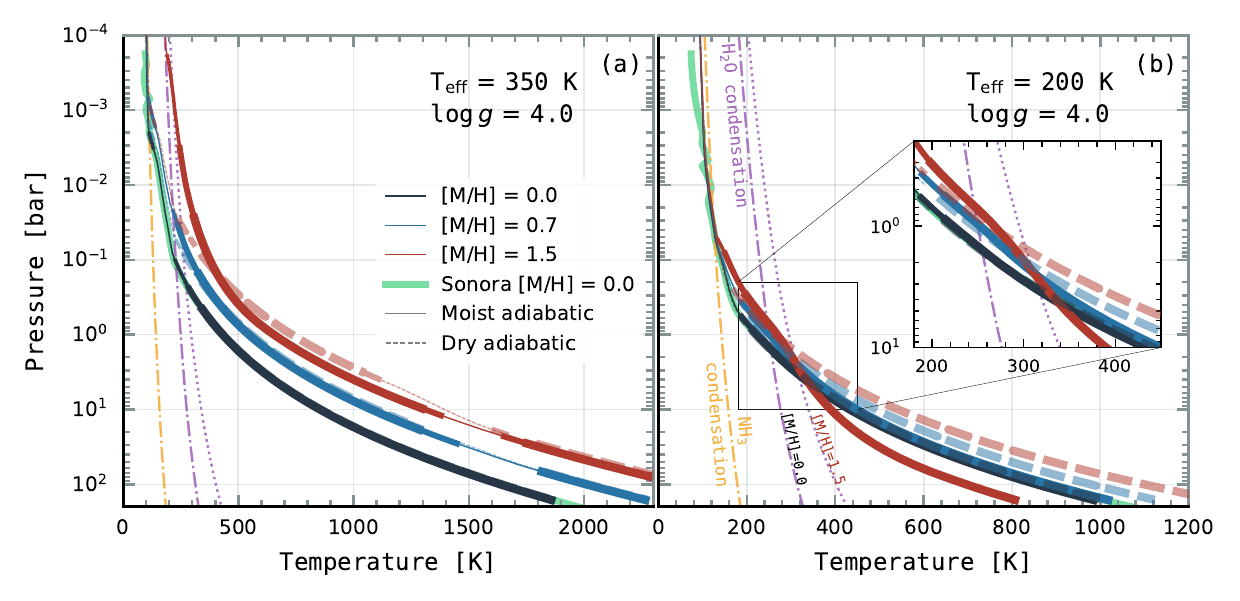}
\caption{
         Thermal structure profiles from converged solutions of the 
         radiative-convective model at \logg = 4.0 for YB ([M/H] of 0.0, here) and YP ([M/H] of 0.7 and 1.5) cases. 
         Panel (a)~shows results from the highest \teff{} investigated (350\,K) and panel (b) shows results from the lowest \teff{} investigated (200\,K).
         Line thickness indicates the fraction of internal heat flux carried by convection; from thin to thick these are:
         0\%, 0\%--1\%, 1\%--10\%, and $>$10\%.
         The Sonora model is from \texttt{Sonora Bobcat} \citep{marl21} 
         with \teff = 200\,K, \logg = 4.0, and [M/H]=0.0.
         Condensation curves at solar metallicity are shown for both water (purple) and ammonia (yellow) as dot-dash lines, and [M/H]=1.5 curves only for water in dotted lines. Impacts of water latent heat release are only apparent at high metallicity for the 350\,K models, and are more pronounced for the 200\,K models (see inset).
         }
\label{fig:pt}
\end{figure*}
%---

\section{Results} \label{sec:results}

The adopted grid of model parameters (Table~\ref{tab:gridpar}) results in a large suite of simulated atmospheric structures and associated spectra. To more clearly demonstrate the impacts of water latent heat, the atmospheric structure and spectral results presented below highlight a limited subset of the modeling grid: effective temperatures of 350\,K, 250\,K, or 200\,K (spanning extremes of the \teff~grid), surface gravity in \logg of 4.0, and metallicities of 0.0, 0.7, and 1.5. Our results related to photometry are more complete in their sampling of the modeling grid. 
Data generated in this study are publicly available\footnote{\dataset[Zenodo: 10.5281/zenodo.5143675]{https://doi.org/10.5281/zenodo.5143675}}.

\subsection{Thermal Structure Profiles}\label{sec:pt}

Figure~\ref{fig:pt} shows equilibrium thermal structure profiles from the aforementioned abridged modeling grid. Solar metallicity models from the \texttt{Sonora Bobcat} grid\footnote{\dataset[Sonora Bobcat Zenodo: 10.5281/zenodo.5063476]{https://doi.org/10.5281/zenodo.5063476}} \citep{marl21} are also shown. The underlying chemistry and opacities for the \texttt{Sonora Bobcat} models are identical to those used here, although the Sonora models use convective adjustment\,---\,rather than mixing length theory\,---\, to treat convective instabilities. Agreement between the \texttt{Sonora Bobcat} models and the solar metallicity simulations developed in this work helps to demonstrate the validity of both grids (and their associated treatments of convection).
In general, effects of latent heat release primarily impact thermal structure gradients near the water condensation curves (which are also depicted in Figure~\ref{fig:pt}). Especially at higher metallicities (implying more atmospheric water vapor), temperature gradients can be markedly reduced due to latent heating.

For \teff~of 350\,K (panel a in Figure~\ref{fig:pt}), atmospheric temperatures remain warm enough that relatively minimal water vapor condensation occurs in the convective region of the atmosphere, independent of metallicity. Regardless of water latent heat treatment, detached convective zones form in the 0.1--10\,bar range for the higher-metallicity models. Detached convective zones were also seen in some solar metallicity Y dwarf models from \citet{morleyetal2014}. These occur when a  window of lower opacity in the deeper atmosphere enables thermal radiative transport to briefly dominate as the preferred mechanism for moving flux through the atmosphere.

At cooler temperatures, as the \teff~of 200\,K models show (panel b in Figure~\ref{fig:pt}), substantial impacts of water latent heat release can be seen, even at a modest metallicity of 0.7 (blue solid line) and especially at a metallicity of 1.5 (red solid line), both of which are YPs in our terminology. The structure of the overlying moist convective region ($\sim$1--10\,bar for the models with \teff~of 200\,K) sets the thermal conditions for the top of the deep, dry convective zone (i.e., below about 10\,bar where H$_2$O is in gas phase). Thus, even the structure of the deep atmosphere is impacted by water latent heat release aloft, which has potential consequences for evolutionary models (see section~\ref{sec:future}).

\begin{figure*}[tb!]
\centering
\includegraphics[angle=0, width=1.\textwidth]{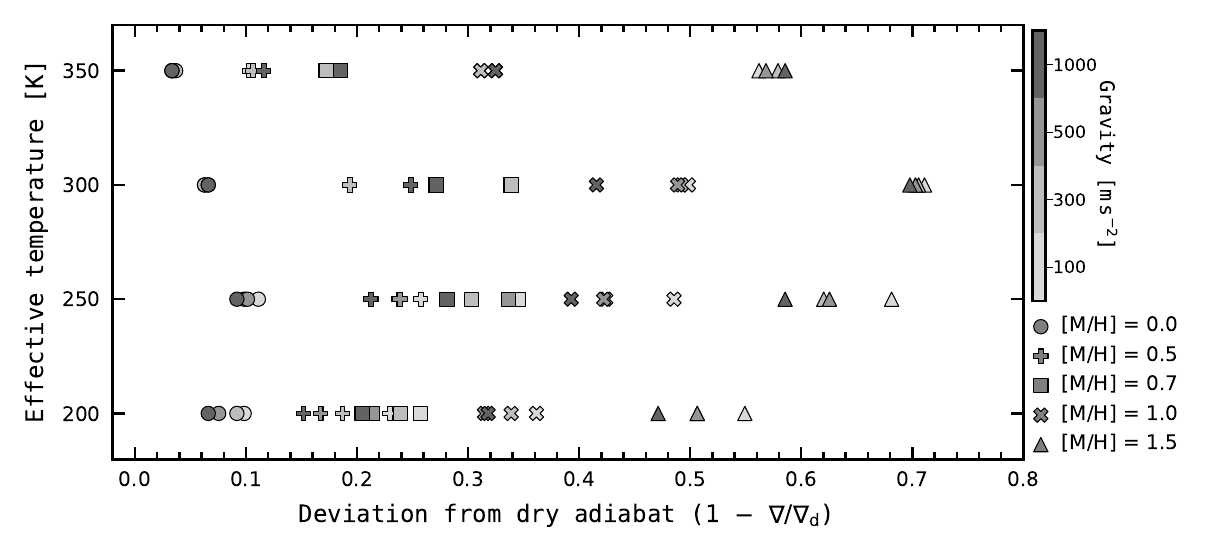}
\caption{
        Most extreme fractional deviation of the moist adiabatic treatment from the dry adiabat (inferred from the minimum value of $\nabla/\nabla_{\rm d}$ achieved in the convective region of the atmosphere) for all simulated YB and YP cases. 
        The minimum deviation for all cases is zero, as shown in Equation~\ref{eq:nabla_m} when no latent heat is present.}
        Metallicity is indicated by symbol type and model gravity is depicted in shades of gray. Deviations are increasingly significant at larger metallicities and lower gravity.	  
\label{fig:gamma}
\end{figure*}

Figure~\ref{fig:gamma} shows the maximum fractional decrease in the moist model adiabat from the dry adiabat for all simulations in the grid. The minimum deviation is zero, which occurs wherever water is not condensing. \citet{legg21} demonstrated that data-model comparisons for some brown dwarfs could be improved through decreasing the adiabatic lapse rate. These authors defined a parameter in terms of the temperature gradient, 
%$\gamma \equiv c_{\rm p}/c_{\rm v} = 1/(1-\nabla)$
$\gamma = 1/(1-\nabla)$, and allowed this parameter to have a non-dry adiabatic value throughout the troposphere while fitting photometry and spectroscopy for several brown dwarfs (at \teff~below  500\,K). Fits indicating that $\gamma \sim$ 1.20--1.33 \cite[Table~3 in][]{legg21} yielded an overall better reproduction of observations (as opposed to using $\gamma \sim$ 1.4, which is more appropriate for a dry adiabat). While water latent heating effects do not impact the entire troposphere, our reported deviations are still useful for comparison to the scale of reductions reported by \citet{legg21}. Values of $\gamma$ for our moist models in convective zone with \teff{}~of 350\,K span $\sim$1.4--1.5 for [M/H] = 0.0 and $\sim$1.1--1.5 for [M/H] = 1.5. As for \teff{}~of 200\,K, $\gamma$ spans $\sim$1.4--1.5 for [M/H] = 0.0 and $\sim$1.2--1.5 for [M/H] = 1.5. Thus, only at YP-like enhanced metallicities can latent heat release provide a partial physical explanation for the lapse rate reductions explored by \citet{legg21}.
%--

\begin{figure}[tb!]
\centering
\includegraphics[angle=0, width=1.\columnwidth]{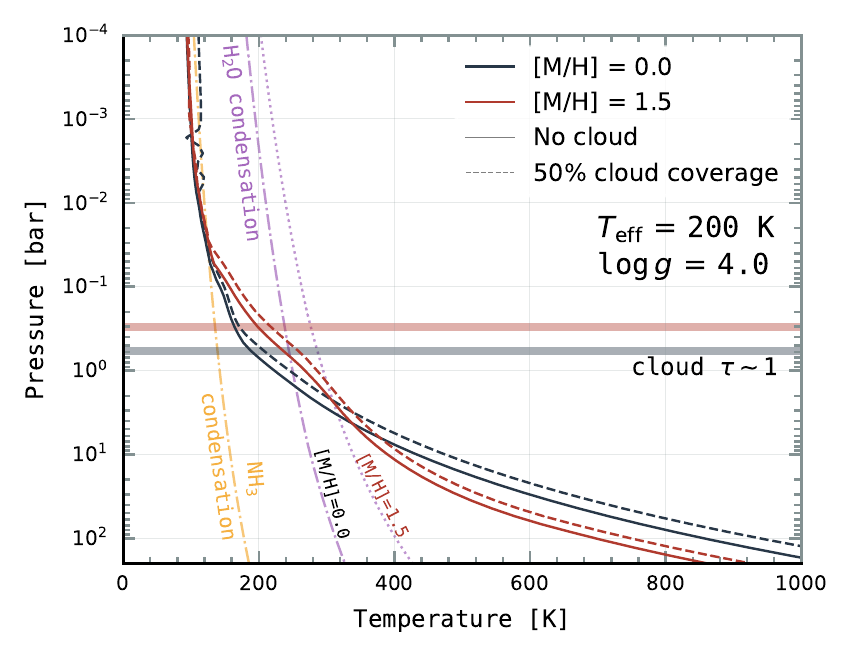}
\caption{
         Example thermal structure solutions with 50\% fractional cloudiness (dotted), as compared to cloud-free models (solid). Models adopt metallicity extremes (a YB with [M/H] of 0.0 [black] and a YP with [M/H] of 1.5 [red]), \teff{} of 200\,K, a \logg of 4.0, and a moist adiabatic treatment. An indicator of the top of the cloud deck (taken as where the cloud column-integrated optical depth achieves unity) are shown as thick horizontal lines.
         }  
\label{fig:cloudPT}
\end{figure}

Figure~\ref{fig:cloudPT} shows the comparison between two sets of temperature profiles from our limited 50\% water cloud coverage models to the cloud-free models. Models shown adopt \teff=200\,K, \logg=4.0, [M/H]=0.0 or 1.5 (i.e., a YB or YP scenario, respectively), and a moist adiabatic treatment. In general, clouds provide an infrared back-warming effect (i.e., greenhouse effect), which results in a warmer deep atmosphere. 

%--

%---
\subsection{Atmosphere Abundances}

\begin{figure}[tb!]
\centering
\includegraphics[angle=0, width=1.\columnwidth]{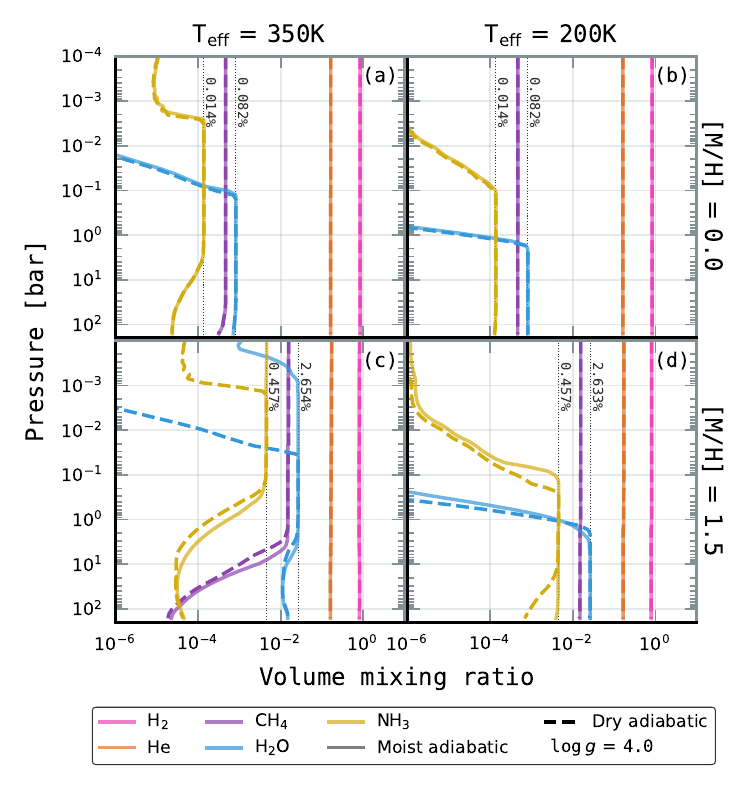}
\caption{
         Chemical abundance profiles for the most abundant species (H$_2$, He, H$_2$O, CH$_4$, and NH$_3$) from converged models. Columns are for different effective temperatures 
         (\teff~of 350\,K and 200\,K at left and right, respectively) and rows are for different metallicities (a YB with [M/H] of 0.0 and a YP with [M/H] of 1.5 at top and bottom, respectively). 
         All models assume \logg = 4.0. 
         Solar metallicity models (panels a and b) have  
         water vapor mixing ratios smaller than 0.1\% while models with [M/H] of 1.5 (panels c and d) have water vapor mixing ratios that can exceed 1\%.
        }	  
\label{fig:mix}
\end{figure}
%---

Figure~\ref{fig:mix} shows converged chemical abundance profiles for four different models. The underlying atmospheric chemistry is determined via an assumption of local rain-out thermochemical equilibrium (Section~\ref{sec:model}). At solar metallicity (panels a and b), the atmosphere is composed primarily ($\gtrsim99.8$\%) of molecular hydrogen and helium, although water vapor can reach a volume mixing ratio of $\sim0.08\%$. Condensation of water can be seen starting near 0.1\,bar and 2\,bar for the 350\,K and 200\,K models, respectively. Even at these relatively low mixing ratios, the impacts of water latent heat release may be detectable (as is discussed in Section~\ref{sec:jwst}). Models at both temperatures also show condensation of ammonia in the upper atmosphere, and the ammonia profile for the 350\,K model indicates a thermochemical preference for nitrogen to exist as N$_2$ in the deep atmosphere.

At enhanced metallicity (Figure~\ref{fig:mix}; panels c and d), the combined molecular hydrogen and helium mixing ratio decreases to roughly $95\%$. The maximum water vapor mixing ratio increases to roughly $\sim 2.6\%$, making it a substantial atmospheric constituent. Interestingly, this water vapor mixing ratio is comparable to surface water vapor concentrations at warmer/moister locations on Earth \citep{mcel02}. As with the solar metallicity models, water vapor and ammonia condensation is apparent. Note that adopting the moist adiabat for the 350\,K model results in a slightly warmer upper troposphere and stratosphere. Thus, substantial water vapor condensation is not seen in the moist case except at very low pressures ($p<10^{-3}$\,bar) in the stratosphere.

%---
\subsection{Model Spectra}\label{sec:spec}

\begin{figure*}[tbh!]
\centering
\includegraphics[angle=0, width=1.\textwidth]{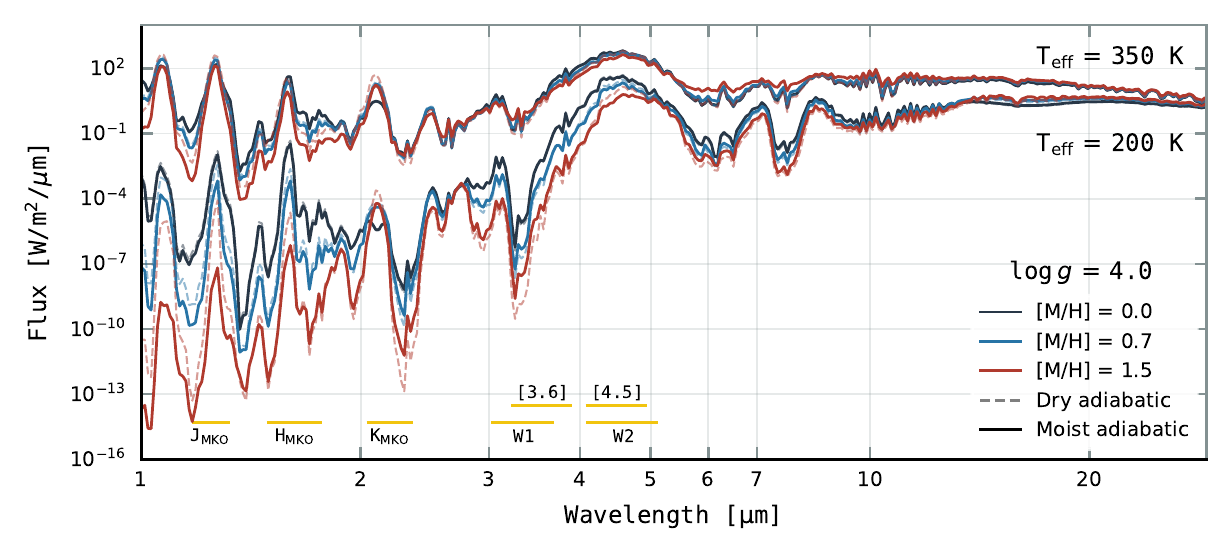}
\caption{
         Low resolution (resolving power of 100) emission spectra for models with \teff~of 350 and 200\,K, [M/H] of 0.0 (YB), 0.7 (YP), and 1.5 (YP), and \logg of 4.0. Spectra assume a size of one Jupiter radius ($R_{\rm Jupiter}$). Photometric bandpasses for Mauna Kea Observatory (MKO) JHK, WISE W1 \& W2, and {\it Spitzer} [3.6] \& [4.5] are indicated.
        }	  
\label{fig:spec}
\end{figure*}
%---

Figure~\ref{fig:spec} shows low-resolving power (R $\equiv \lambda/\Delta \lambda=$ 100) emission spectra for previously-shown thermal structure models (Figure~\ref{fig:pt}) generated with the \texttt{PICASO} open source software\footnote{\url{https://natashabatalha.github.io/picaso/}.}  \citep{bata19,bata20picasoZenodo}. As before, adopted models bracket the effective temperatures explored here (i.e., 350\,K and 200\,K) and assume \logg of 4. Three different metallicities are shown (with [M/H] of 0.0, 0.7, and 1.5), as are the moist adiabatic versus dry adiabatic cases. 

At the lower effective temperature (200\,K), flux differences for the varied metallicities are most strongly distinguished at near-infrared wavelengths. A difference in flux of roughly a factor of $10^6$ is shown in the J band between metallicities of 0.0 and 1.5. Moreover, the difference in flux between the dry and moist treatments is also most distinct in this wavelength region. Nevertheless, with emitted flux peaking at longer near-infrared and mid-infrared wavelengths, it may be that differences in models may be most straightforward to observe at the longer wavelengths where greater photon fluxes are achieved.

\begin{figure*}[tbh!]
\centering
\includegraphics[angle=0, width=1.\textwidth]{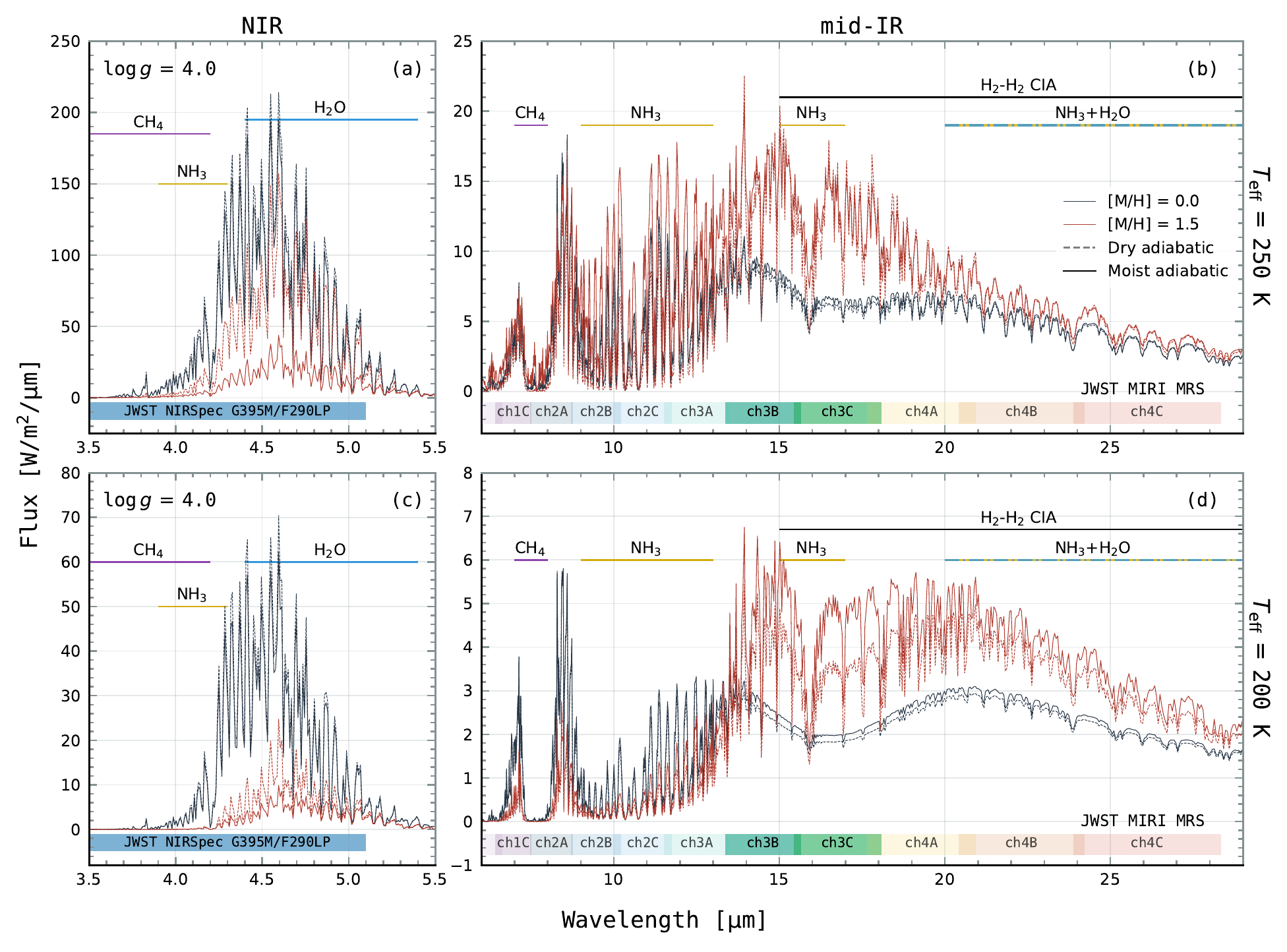}
\caption{
         Model spectra (resolving power of 500) in near-infrared
         (NIR; panels a and c) and mid-infrared (mid-IR; panels b and d) spectral regions for \teff~of 250 and 200\,K (top and bottom rows, respectively) with \logg = 4.0. Both a YB (black) and YP (red) case are shown. The {\it JWST} MIRI MRS filter regions are shown at the bottom of (b) and (d). Key opacity regions for CH$_4$, H$_2$O, NH$_3$, and H$_2$-H$_2$ collision-induced absorption (CIA) are indicated.
        }	  
\label{fig:spec250}
\end{figure*}

\begin{figure*}[tbh!]
\centering
\includegraphics[angle=0, width=0.95\textwidth]{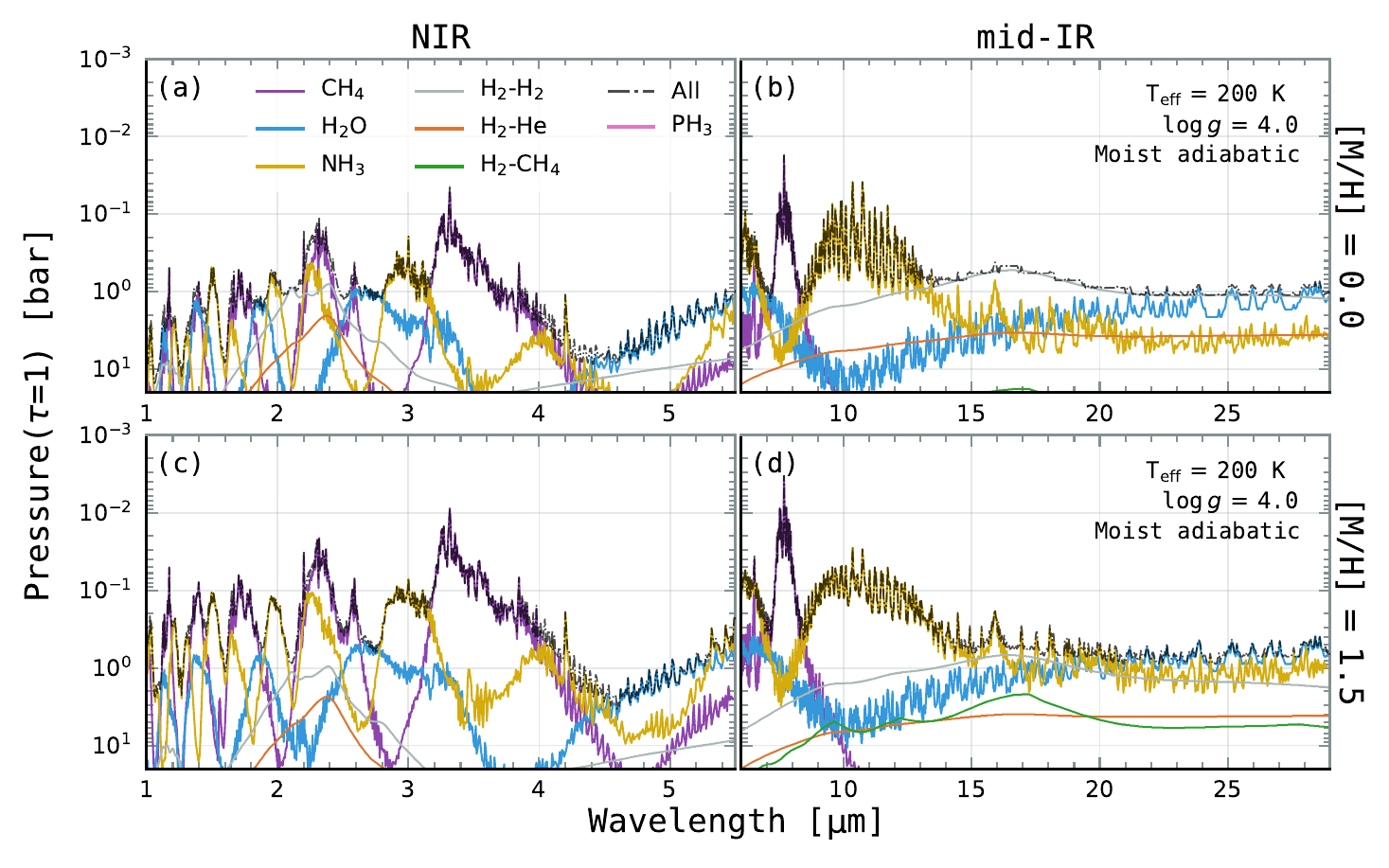}
\caption{
         Wavelength-dependent location of the ``photosphere'' (i.e., the atmospheric pressure level where column-integrated optical depth, $\tau$, reaches unity) and contributions from key opacity sources. Left columns show the near-infrared wavelength range and right columns show the mid-infrared wavelength range. All plots adopt \teff = 200\,K, \logg = 4.0, and a moist adiabat. Upper rows show [M/H] of 0.0 (YB) while bottom rows show [M/H] of 1.5 (YP).
        }	  
\label{fig:tau}
\end{figure*}

Figure~\ref{fig:spec250} shows higher-resolution ($R=500$) spectra spanning limited wavelength ranges in both the near-infrared (3.5--5.5~$\upmu$m) and mid-infrared (6--29~$\upmu$m). More than 99\% of the flux for these models is emitted in the in 1--30~$\upmu$m range. The upper panels (a and b, with \teff~of 200\,K and \logg of 4.0) represent an analog for the nearest ($\sim$2.3 pc) and coolest Y-dwarf known to date\,---\,WISE 0855$-$0714, with \logg$\sim$3.5--4.3, \teff$\sim$240--260\,K, and mass$\sim$1.5--8.0\,$M_{\rm Jupiter}$ \citep{legg17}. Figure~\ref{fig:tau} helps to indicate key opacity sources by showing contributions at the level of the photosphere (here, taken as the wavelength-dependent atmospheric level where the column-integrated optical depth achieves unity). In the highlighted near-infrared wavelength region, the dominant opacity sources at shorter wavelengths are methane and ammonia while water opacity dominates at redder wavelengths. Strong opacity contributions from ammonia, water, and H$_2$-H$_2$ collision-induced absorption (CIA) are apparent in the highlighted mid-infrared wavelength range, especially at wavelengths longer than 16\,$\upmu$m.

%---

%---

\begin{figure*}[tb!]
\centering
\includegraphics[angle=0, width=1.\textwidth]{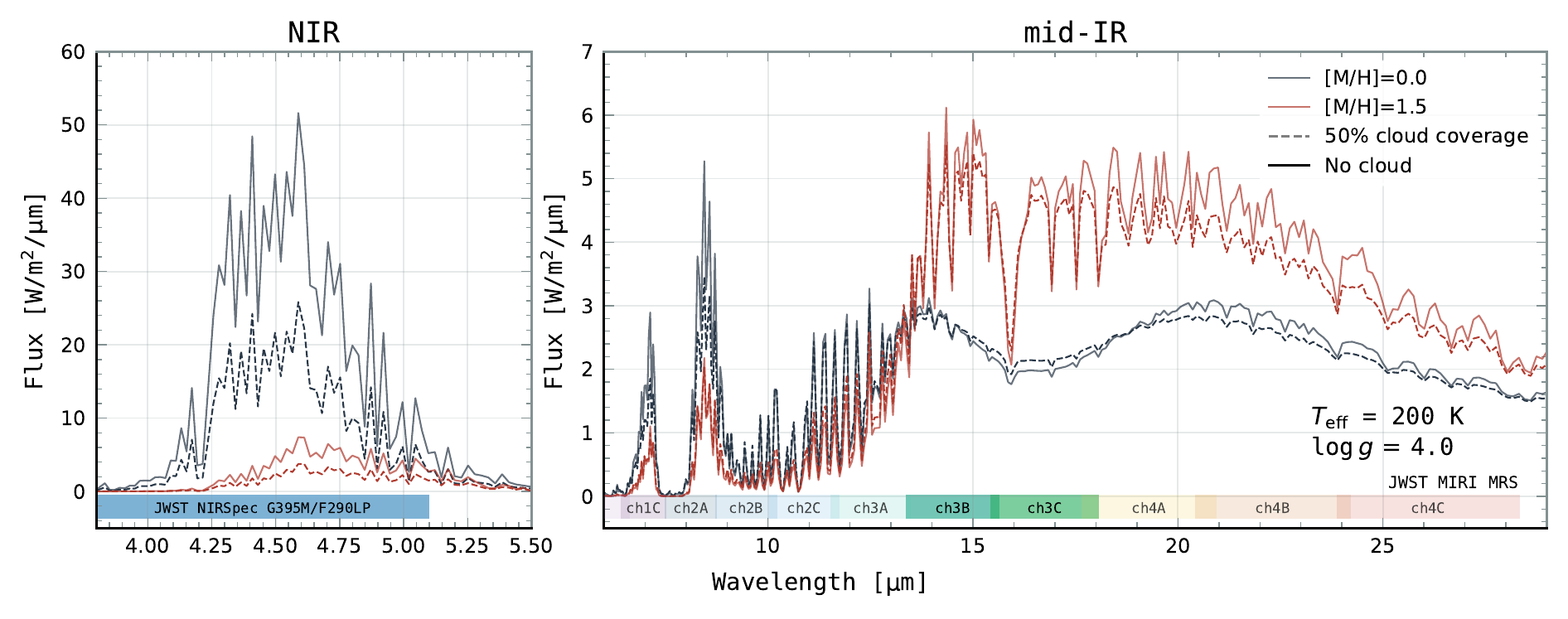}
\caption{
         Emission spectra (resolving power of 200) comparison of cloud-free and 50\% water cloud coverage models. Models shown are for \teff= 200\,K, \logg=4.0 and [M/H] of 0.0 (YB) and 1.5 (YP). Spectra assume a size of one Jupiter radius ($R_{\rm J}$). Photometric bandpasses for {\it JWST} shown are the same as in Figure~\ref{fig:spec250}. 
        }	  
\label{fig:cloud_spec}
\end{figure*}

Figure~\ref{fig:cloud_spec} (similar to Figure~\ref{fig:spec}) shows spectral comparisons between cloud-free and 50\% water cloud coverage models with [M/H] of 0.0 (YB) or 1.5 (YP) and \logg of 4.0. The spectral region affected most by clouds is the near-infrared wavelength region ($\sim$4.5 $\rm \upmu m$), which is a gas opacity window (Figure~\ref{fig:tau}). Here, flux can more easily escape from the clearsky model than the cloudy model (where water cloud opacity partially obscures the window). Owing to generally warmer thermal structures (Figure~\ref{fig:cloudPT}), mid-infrared fluxes from cloudy models can exceed those from cloud-free solutions.

%---
\subsection{Color-Magnitude Diagrams}

\begin{figure*}[tb!]
\centering
\includegraphics[angle=0, width=1.\textwidth]{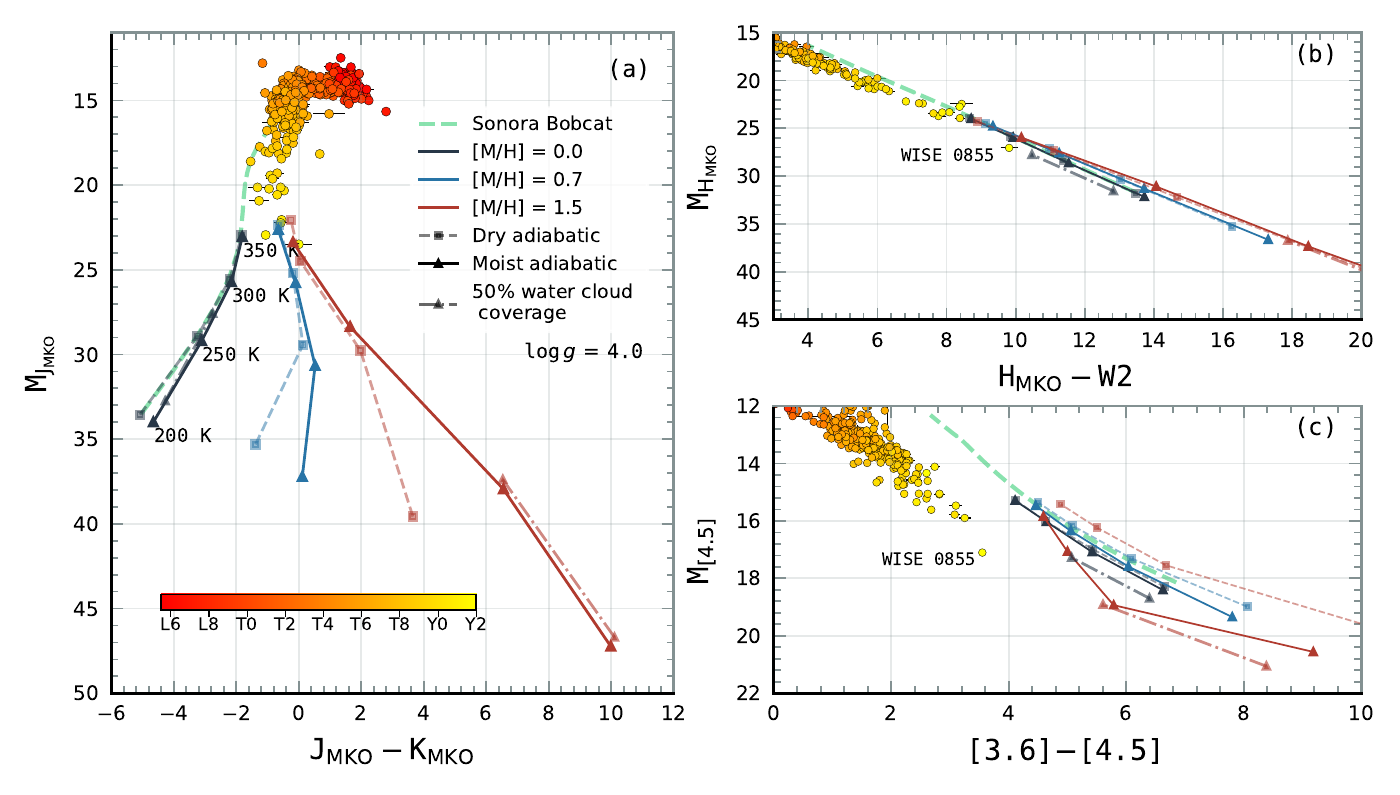}
\caption{
         Color-magnitude diagrams for model grids compared to observational data from \citet{best20}. Data are for field brown dwarfs with
         no known unresolved companions from L6 to Y2 spectral types. Moist (triangle/solid) and dry (square/dashed) model tracks are shown for metallicities of 0.0 (YB), 0.7 (YP), and 1.5 (YP; black, blue, and red, respectively). Solar metallicity, cloud-free models from the \texttt{Sonora Bobcat} grid (green dashed) are shown for validation purposes. Finally, a limited number of 50\% cloudy runs (dash-dot) are provided at high metallicity to investigate the impacts of water clouds.
        }	  
\label{fig:cmd}
\end{figure*}

To better facilitate data-model comparisons, Figure~\ref{fig:cmd} shows model photometry (assuming one Jupiter radius) in filters from the Mauna Kea Observatory (MKO) photometry system, the Wide-field Infrared Survey Explorer (WISE), and the {\it Spitzer Space Telescope}\footnote{Filter profiles and Vega magnitude zero points are from the SVO Filter Profile Service (\url{http://svo2.cab.inta-csic.es/theory/fps/})} (see also Appendix Table~\ref{tab:phto}). The data in Figure~\ref{fig:cmd} are from \citet{best20}, and are for isolated field brown dwarfs. Solar metallicity, dry adiabatic model results from the \texttt{Sonora Bobcat} grid \citep{marl21} are also shown, and indicate good agreement with the equivalent simulations performed here. A limited number of models that include partial water cloud coverage (50\%) \citep[following the treatments of][]{morleyetal2014} and moist convection are plotted to help illustrate the limited impact of clouds on the near-infrared spectral region.

The J versus J$-$K color-magnitude diagram (Figure~\ref{fig:cmd} panel a) shows the standard shifts towards bluer colors at the L-T transition, followed by a slight reddening for Y dwarfs. Cloud-free, solar metallicity models (i.e., YB models) do not reproduce the Y dwarf color trends.
While YP models (with enhanced metallicity) do reproduce the limited number of Y dwarf observations, this explanation is unsatisfying as it requires all of these field objects to have large metallicities. Data-model mismatches for YBs can be a result of imperfect modeling of J band gas opacity \citep[see][]{marl21}, where enhancements of CH$_4$ and/or NH$_3$ may help to drive the J$-$K color redder.
In contrast, the H versus H$-$W2 color-magnitude diagram shows the tight linear trend for ultra-cool objects first described in \citet{kirk20}. Based on the model grid developed here, this linear trend\,---\,an important tool to identify unresolved companions\,---\,could continue down to (at least) \teff~of 200\,K. Finally, the {\it Spitzer} [4.5] versus [3.6]$-$[4.5] color-magnitude diagram shows a significant difference between the data and models. Such discrepancies are a known problem \citep{legg17} and discussed in Section~\ref{sec:future}.

%---------------------------------------------------------------
%%%%%%%%%%%%%%%%%%%%%%%%%%%%%%%%%%%%%%%%%%%%%%%%%%%%%%%%%%%%%%%%%%%%%%%%%%
%---------------------------------------------------------------

\subsection{Simulated James Webb Space Telescope Observations}\label{sec:jwst}

The 6.5 meter {\it JWST} is a joint NASA-ESA-CSA mission scheduled to launch in late 2021 \citep{gardneretal2006}. The near- and mid-infrared capabilities of {\it JWST} make it well-suited to the study of ultra-cool brown dwarfs and free-floating planets. The broad wavelength coverage of the Near-Infrared Spectrograph \citep[NIRSpec, 0.6--5.3\,$\upmu$m;][]{bagn07} and the Mid-Infrared Instrument \citep[MIRI, 4.9--28.3\,$\upmu$m;][]{riek15} make them particularly well-suited to observing and characterizing Y dwarfs. While the exposure times presented here emphasize NIRSpec, Figure~\ref{fig:spec250} indicates that the mid-infrared wavelength range spanned by MIRI includes indicators that are sensitive to the metallicity of a given Y dwarf.

The NIRSpec instrument provides several different modes: Multi-Object Spectroscopy (MOS), imaging spectroscopy with the Integral Field Unit (IFU), high contrast single object spectroscopy with the Fixed Slits (FSs), and high throughput Bright Object Time-Series (BOTS) spectroscopy. The FSs mode with the G395M disperser and the F290LP filter is well-suited to study the 4.5\,$\upmu$m window (highlighted in Figure~\ref{fig:spec250}), as this disperser/filter combination spans 2.87--5.10\,$\upmu$m with $R \sim 1000$ resolving power.
Given the faintness of potential targets explored here, and hence the possibilities of blind guiding, the S400A1 aperture with its larger slit width of $\sim$0.4\arcsec is adopted. The \texttt{Pandeia} {\it JWST} observation simulation tool  \citep{pont16}\footnote{\url{https://pypi.org/project/pandeia.engine/##description}} was used for estimating signal-to-noise ratios and exposure times. In \texttt{Pandeia}, the signal to noise (S/N) is estimated based on the number of groups in each ramp, the number of ramps in each exposure, and the number of user-specified exposures.

\begin{figure}[tbh!]
\centering
\includegraphics[angle=0, width=1.0\columnwidth]{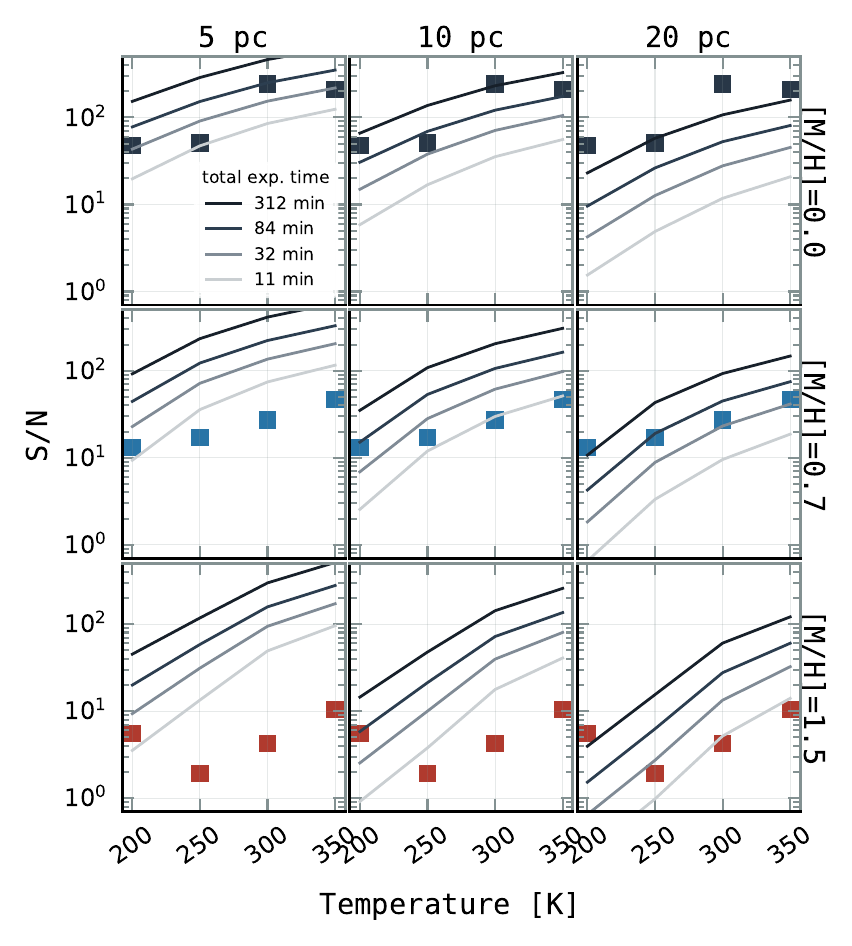}
\caption{
         Detectability of the impacts of moist convection for models with different effective temperatures, at three different distances (5, 10, and 20\,pc at left, middle, and right, respectively), and a \logg~of 4. Square symbols show the observational S/N needed to detect the spectral impact of moist convection at a 5$\sigma$ level in the 4.5\,$\upmu$m window region. Lines indicate the S/N delivered by {\it JWST} NIRSpec G395M/F290LP in different exposure times (312, 84, 32, and 11 minutes; gray shades), as estimated by the \texttt{Pandeia} exposure time calculator. Thus, cases where gray lines sit above square datapoints indicate stronger detectability. Exposure times assume a single ramp in each exposure, and the combination of the number of exposures and the number of groups in each ramp are (30, 100), (20, 40), (15, 20), and (10, 10) for the 312, 84, 32, and 11-minute cases, respectively.
        }	  
\label{fig:jwstSN}
\end{figure}

Evidence for moist convection is quantified using the flux difference between a dry versus moist adiabatic model at a given effective temperature, metallicity, and gravity. With this definition, Figure~\ref{fig:jwstSN} shows the detectability (at a S/N of 5) of the impacts of moist convection for various objects at two different distances (5 and 10 parsecs) and for three different exposure times. Squares indicate the requisite S/N for detection while lines indicate S/N values from \texttt{Pandeia}, implying that the impacts of moist convection could be detected in cases where a line sits above a square datapoint. The best scenario explored here occurs when the target is located at a distance of 5\,pc and with metallicity of 1.5, i.e., a close-by free-floating planet. Here the moist convection effects at all \teff{} can be detected with about 30 minutes of exposure. On the other hand, the most difficult scenario explored here occurs when the target located at 10\,pc and has solar metallicity. In this case, an exposure time longer than 1.5 hours would be required to detect the impacts of moist convection.

%---

%----------
\section{Discussion} \label{sec:disc}

The results presented above have several important consequences for the characterization of Y-class brown dwarfs and free-floating exoplanets. These consequences are best understood through the impacts of metallicity and moist convection on observables, which is discussed immediately below. However, models still fall short of reproducing some observations, so areas for model improvements are also discussed.

%----------
\subsection{Detecting Effects of Metallicity and Moist Convection}

There are two primary models for how giant planets are distinguished from brown dwarfs: based on a minimum mass for deuterium burning or based on formation mechanism. Characterizations based on the deuterium burning limit may be practical, but are influenced by model dependencies \citep{grossman1970,grossmanetal1974,burrowsetal1997,spie11,bode13} and associated issues with constraining masses from observations \citep{chau05,curr14}. As an example of the subtleties involved, a 13 Jupiter-mass object could be classified as a planet or a brown dwarf depending on its initial helium, deuterium, and metal abundances \citep{spie11}.

Categorization based on formation mechanism (core accretion versus gravitational collapse) is physically and observationally well-motivated \citep{Burrows01,chab14,schl18}. While such a distinction may be more straightforwardly applied to worlds orbiting stellar hosts where the formation history is likely much better-known, issues could arise when studying sub-stellar objects in the field. Here, distinguishing a free-floating planet\,---\,which would have been ejected from a planetary system following formation\,---\,from a brown dwarf would include an emphasis on the metallicity of the object, as formation via core accretion \citep{poll96} can lead to much higher levels of metals enrichment. Thus, the metallicity-dependent models presented above can help interpret observations of Y dwarf targets as either free-floating planets or brown dwarfs (although cases with intermediate metallicities will present challenges to interpretation). Furthermore, as the models described above indicate that the impacts of moist convection are sensitive to metallicity, observables related to water latent heat release can also help distinguish free-floating planets from brown dwarfs.

Figure~\ref{fig:spec250} shows that the 4.5\,$\upmu$m spectral window could be a powerful tool for indicating metallicity in ultra-cool objects and, thus, their formation mechanism. For the 250\,K, case, the moist adiabatic models show a $\sim$80\% flux difference between a solar metallicity and a metal-rich object. Dry models show only a $\sim$40\% flux difference, which demonstrates that water latent heating effects are just as important as underlying chemical abundances in influencing fluxes in this spectra range. Assuming targets in the effective temperature range of 200--350\,K are well-represented by moist models.

Figure~\ref{fig:jwstSN} indicates that objects with [M/H] spanning 0.0 to 1.5 (i.e., YB through YP objects) within 10\,pc may be distinguished spectroscopically at the 5$\sigma$ level with {\it JWST} exposure times of less than about 1.5 hours for nearly all effective temperatures in our grid. While a 5$\sigma$ detection could still be achievable within about 1.5 hours of exposure time for high metallicity (YP) cases at 20\,pc, detections at solar metallicity (i.e., YB-like scenarios) become challenging. 
Atmospheric retrieval as applied to high-quality brown dwarf spectral observations will also constrain metallicity \citep{lineetal2014b,burninghametal2017,lothringer&barman2020,piette&madhusudhan2020} and such inference tools may need to be updated to allow for detections of markedly sub-adiabatic temperature gradients.

Ammonia is also likely to be an important tracer of metallicity, as indicated by near- and mid-infrared spectra in Figure~\ref{fig:spec250}. Shifting of the underlying continuum makes the ammonia feature near 4.2\,$\upmu$m much less pronounced for higher-metallicity models. In the mid-infrared, a strong, broad ammonia feature near 16\,$\upmu$m is overwhelmed by molecular hydrogen CIA for lower-metallicity models. Observations at these two wavelengths, even if only photometric, could efficiently trace Y dwarf metallicity.

While the models presented here are largely cloud-free, previous Y dwarf modeling studies have shown somewhat limited impacts of water clouds (at solar metallicity), especially in the 4.5\,$\upmu$m spectral window \citep[][their Figure~11]{morleyetal2014}. Spectra in this window probe the deeper atmosphere (below roughly 1\,bar), and large fluxes from this warmer portion of the atmosphere can dominate over the more-limited fluxes emerging from patches of the atmosphere obscured by cooler clouds. Nevertheless, cloudiness remains an important consideration when interpreting observations of ultra-cool objects \citep{skemeretal2016,morleyetal2018}.

%----------
\subsection{Future Work and Model Improvements} \label{sec:future}

While results indicate that moist convection is likely to have an observable impact through a swath of the Y spectral class with \teff $\lesssim$350\,K, data-model comparisons still show room for improvements. Most significantly, {\it Spitzer} [4.5] versus [3.6]$-$[4.5] color-magnitude diagrams (Figure~\ref{fig:cmd} show systematic discrepancies). Poor model fits in these bands are a known issue \citep{kirk19}, where disequilibrium chemistry, breaking gravity waves, and clouds have been discussed as potential solutions \citep{morl18}. At the low temperatures investigated here, a disequilibrium chemistry mechanism for carbon would not apply as carbon is stable in the form of CH$_4$. Furthermore, \citet{morleyetal2014} demonstrated only minimal impacts of disequilibrium chemistry at effective temperatures below about 300\,K. Breaking gravity waves may be a viable explanation, but would be required to produce colors that are roughly 1 magnitude bluer than those presented here, which is a significantly larger effect than has been investigated with previous models that include upper-atmospheric heating \citep{morl18}. As in the case study of \citet{morl18}, it may be that a combination of the effects of metallicity, carbon-to-oxygen ratio, disequilibrium chemistry, and clouds are all required to produce acceptable fits.

Models presented here would benefit from a more physical treatment of convection and associated cloud formation as well as pairing to evolutionary calculations. Convection is, by necessity, parameterized in one-dimensional models, and is influenced by competing effects of both temperature and mean molecular weight gradients in atmospheres \citep{tremblinetal2015,tremblinetal2016}. Rather than relaxing thermally unstable regions of the atmosphere to the moist adiabat, improved dynamic solutions would only trigger (dry) convection once an atmospheric layer becomes net buoyant and would only release latent heat (and form clouds) once appropriate conditions for condensation have been met. Interestingly, models presented here also show that ammonia could be condensing in the convectively-stable stratospheres of some Y dwarfs, where an associated ammonia volatile cycle would be dependent on stratospheric mixing and dynamics \citep{showman&kaspi2013}. Finally, as water latent heat release impacts the deep thermal structure of a wide range of ultra-cool models presented here, evolutionary calculations \citep[which rely on the deep atmospheric adiabat as a boundary condition; e.g.,][]{saumon&marley2008} should be updated at low effective temperatures.

%---------------------------------------------------------------
%%%%%%%%%%%%%%%%%%%%%%%%%%%%%%%%%%%%%%%%%%%%%%%%%%%%%%%%%%%%%%%%%%%%%%%%%%
%---------------------------------------------------------------
\section{Conclusions} \label{sec:conc}

Work presented here investigated the impact of water latent heat release on the atmospheric thermal structures of Y-class brown dwarfs and free-floating planets. Key results are as follows:
\begin{itemize}
    \item The one-dimensional, dynamic atmospheric structure model for brown dwarfs and giant exoplanets developed in \citet{robinson&marley2014} has been updated to include moist convective effects.
    \item A grid of atmospheric simulations spanning a range of metallicities, gravities, and effective temperatures shows that water latent heating effects are likely to be significant at effective temperatures below about 350\,K, especially at super-solar metallicities. At high metallicities ([M/H] of, or above, 1.5), water vapor mixing ratios can reach an Earth-like 3\% and latent heating effects are pronounced.
    \item Spectral impacts of metal enrichment and moist convection will influence near- and mid-infrared observations of Y dwarfs. Such impacts could be straightforward to detect with {\it JWST} for Y dwarfs at distances of up to 10\,pc (or beyond).
\end{itemize}

%-----------------------------------
%-----------------------------------
%-----------------------------------

\acknowledgments

We express our gratitude to an anonymous referee for providing comments and suggestions that helped to improve the quality of this paper.
Work presented here made use of the SVO Filter Profile Service (\url{http://svo2.cab.inta-csic.es/theory/fps/}) supported from the Spanish MINECO through grant AYA2017-84089.
This research also benefited from The UltracoolSheet, maintained by Will Best, Trent Dupuy, Michael Liu, Rob Siverd, and Zhoujian Zhang, and developed from compilations by \citep{dupu12,dupu13,liu16,best18,best21}. TDR gratefully acknowledges support from NASA's Exoplanets Research Program (No.~80NSSC18K0349), Habitable Worlds Program (No.~80NSSC20K0226), and Nexus for Exoplanet System Science and NASA Astrobiology Institute Virtual Planetary Laboratory (No.~80NSSC18K0829).

\vspace{5mm}

\software{\texttt{astropy} \citep{astropy18},  
          \texttt{PICASO} \citep{bata19,bata20picasoZenodo},
          \texttt{Pandeia} \citep{pont16}
          }

%-----------------------------------
%-----------------------------------
\clearpage
\appendix{}
\section{Synthetic Photometry}\label{sec:app}

Synthetic photometry transfer from model spectra with R $\sim$3000. 
%-----------------------------------
\startlongtable
\begin{deluxetable}{ccccccccccccc}
\tablecaption{Synthetic photometry\label{tab:phto}
              }
\tabletypesize{\footnotesize}
\tablehead{ 
 	 \colhead{$g$}    & \colhead{\teff}   & \colhead{[M/H]}  &
 	 \colhead{IRAC36}   & \colhead{IRAC45}  & \colhead{W1}      & \colhead{W2}      &
 	 \colhead{GiminiM}  & \colhead{MKO\_J}   & \colhead{MKO\_H}   & \colhead{MKO\_K}   &
 	 \colhead{MKO\_Lp}   & \colhead{adiabatic$^\dagger$ }\\
 	 \colhead{(\cmst)}    & \colhead{(K)}     & \colhead{      }  &
 	 \multicolumn{9}{c}{(mag)}  & \colhead{  }\\
 	 \cline{4-12}
 	 \colhead{(1)} & \colhead{(2)} & \colhead{(3)} & \colhead{(4)} & \colhead{(5)} & \colhead{(6)} & \colhead{(7)} &
	 \colhead{(8)} & \colhead{(9)} & \colhead{(10)} & \colhead{(11)} & \colhead{(12)} & \colhead{(13)}
 	 }
\startdata
100 & 200 & 0.0 & 24.94 & 18.29 & 26.40 & 18.32 & 17.76 & 33.56 & 31.78 & 38.64 & 22.61 & D \\
100 & 200 & 0.0 & 25.03 & 18.41 & 26.47 & 18.43 & 17.87 & 33.99 & 32.15 & 38.64 & 22.72 & M \\
100 & 250 & 0.0 & 22.41 & 16.98 & 23.70 & 16.99 & 16.48 & 28.91 & 28.36 & 32.15 & 20.52 & D \\
100 & 250 & 0.0 & 22.51 & 17.09 & 23.79 & 17.09 & 16.58 & 29.21 & 28.63 & 32.31 & 20.63 & M \\
100 & 300 & 0.0 & 20.64 & 16.01 & 21.77 & 16.00 & 15.54 & 25.58 & 25.84 & 27.79 & 19.00 & D \\
100 & 300 & 0.0 & 20.68 & 16.03 & 21.82 & 16.03 & 15.56 & 25.72 & 25.95 & 27.88 & 19.04 & M \\
100 & 350 & 0.0 & 19.38 & 15.27 & 20.40 & 15.26 & 14.82 & 22.99 & 23.92 & 24.82 & 17.89 & D \\
100 & 350 & 0.0 & 19.40 & 15.29 & 20.43 & 15.28 & 14.85 & 23.06 & 23.99 & 24.87 & 17.92 & M 
\enddata
\tablecomments{
    $^\dagger$ ``D'' for dry adiabatic model and ``M'' for moist adiabatic model.
    This table is available in its entirety in machine-readable from in the online journal.
    }
\end{deluxetable}

%-----------------------------------
%-----------------------------------
\bibliography{main}
\bibliographystyle{aasjournal}

% \begin{thebibliography}{}
%\bibitem[Leggett et al.(2021)]{legg21} 
%        {Leggett}, S.~K. et al. \ 2021, \apj, in press
        
% \bibitem[Marley et al.(2021 submitted)]{marl21} 
%         {Marley}, M. et al. \ 2021, \apj, submitted

% \end{thebibliography}{}
%-----------------------------------
%-----------------------------------

\end{document}